# Evolution of liquid crystal microstructure during shape memory sequence in two main-chain polydomain smectic-C elastomers

Sonal Dey[1,*], XX[†], XX[#], XX[2], and XX[2]

*[1] Department of Physics, Kent State University, Kent, OH 44242*
*[2] XX*

Two main-chain smectic-C elastomers were investigated with synchrotron x-ray diffraction to understand the evolution of liquid crystal (LC) microstructure under external strain and its relationship to soft-elasticity and shape-memory effect. The experiments reveal the presence of two different relaxation mechanisms in these systems at low and high strains. At low strains, the smectic layers are reoriented with layer-normals distributed in a plane perpendicular to the stretch direction. The system relaxes relatively slowly (time-constant ~ 45 minutes) which is attributed to the flow properties of the LC layers embedded in the elastomer network. At high strains, the equilibration time (~ 4 - 8 minutes) conforms to a faster relaxation and appears to have its origin in the polymer components of the system. Due to misaligned microdomains at small strains, the value of global orientational order parameter $S$ for the mesogenic parts is initially small (~ 0.15). With increasing strain, the local domain-directors, the mesogens, and the polymer chains, all tend to align parallel to the stretch direction giving rise to a higher measured value of $S$ ~ 0.83 at a strain of 4.0. The siloxane segments remain less ordered, attaining a value of $S$ ~ 0.4 for a strain of 4.0. The layers gradually become oblique to the stretch direction conforming to the structural property of the smectic-C phase. The system finally assumes a chevron-like optically monodomain structure. Both elastomers are "locked-in" this state even after removal of the external stress giving rise to strain retention and the shape memory effect. The presence of a transverse component in the main-chain leads to higher strain retention in the second elastomer. A preference for the orientation of the smectic layer-normals toward the stretch direction persists after removal of external stress. Upon thermal annealing, the chevron-like microstructure gradually melts via a different path to the initial polydomain structure.

**PACS:** 61.41.+e , 61.30.-v, 61.05.cf, 83.80.Va

[*] Current address: Department of Physics, Astronomy and Materials Science, Missouri State University, Springfield, MO 65897. Correspondence to:sdey@kent.edu
[†] XX
[#] XX

## I. INTRODUCTION

A synthetic elastomer is typically composed of cross-linked network of polyisoprene [1], devoid of any microstructure [2,3]. The network can be reversibly deformed more than five times the initial length and has low values for the elastic modulus [1,4]. Such large scale reversible deformation is usually explained by assuming a Gaussian distribution of the polymer segments between the cross-linking points [1,5,6]. But the Gaussian approximation is not adequate in explaining the upward turn [1] in the stress-strain curve at very high extensions. At this point, the effect of finite extensibility [7] of the network comes into picture and the chain statistics is approximated by the inverse Langevin function [8]. The first order approximation of the inverse Langevin function reduces exactly to the case of the Gaussian chain [1,8]. Thus, for all practical purposes, the polymer chain conformation could be assumed to be Gaussian [1].

Introduction of liquid crystalline (LC) components into the elastomer network modifies the shape of the polymer chain [9]. The chain statistics is now described by an anisotropic Gaussian distribution in the purview of the neo-classical elastic theory [9] coined by Warner and Terentjev. The step-length of the polymer chain now becomes a tensorial quantity [9] and the elongation of the network becomes highly direction dependent. The natural tendency of the polymeric network is to disorder and maximize entropy [6]. This now combines with the orientational and/or translation order of the LC molecules [10] leading to many interesting effects, *e.g.*, large-scale thermo-mechanical actuation [11-15], soft elasticity [9,16-18] and shape memory effect [19-22].

The soft-elastic response of the LC elastomer network is seen in that portion of the stress-strain curve where a large deformation could be achieved without any significant rise in the value of stress [9]. The deformation seems to take place with almost zero resistance in that region. The phenomenon can be explained based on a soft-mode of director rotation of the mesogens [16,17,23] that leaves the elastic free energy invariant. Physically, in case of an initially polydomain elastomer [24,25], the LC components assemble into small domains. The orientations of the domains are initially random such that light is scattered in different directions which makes the elastomer specimen opaque to the incoming light [9]. As one applies strain, the domains which are not parallel to the stretch direction, gradually orient toward it and the elastomer eventually becomes an optically homogeneous medium for the incident light [9]. This shear-induced re-orientation of the individual liquid crystalline directors [23,26,27] leads to a plateau in the stress-strain curve. In this region, the elastomer elongates with negligible increase in stress. The stress will again rise when this re-orientation is completed. In case of a nematic elastomer, removal of the external load brings the system back to the initial polydomain state [9,24].

In case of initially polydomain smectic elastomers, smectic micro-domains can form inside the polymer network [28,29]. The introduction of quasi long-range positional order [10] could lead to significant strain-retention [30] which is the basis of shape memory effect [20] in these systems. From the view point of energy, the layers will prefer to rotate [31] in response to an applied strain rather than being deformed. This is because the smectic elastic modulus is typically two orders of magnitude higher than the rubber elastic modulus [32,33]. Such rotation of the smectic layers inside the elastomer network amounts to a polydomain to monodomain transition [28,30,34-38] in these systems.

Sánchez-Ferrer*, et al.,* [35] reported the polydomain to monodomain transition caused by uniaxial strain in main-chain smectic-C elastomers. The monodomain obtained is termed *pseudo-monodomain* [39] because of the conical distribution of the layers [40] around the mechanically induced director. For brevity, we shall use the term *monodomain* to imply sample with a well-defined direction for the director, which accompanies conical distribution of smectic layers around the director.

In this study, two main-chain smectic-C elastomers (initially in the polydomain state) have been investigated with synchrotron x-ray diffraction to understand the evolution of LC microstructure under applied strain and its relationship to soft-elasticity and the shape memory effect. The elastomers were subjected to uniaxial strain, allowed to relax at constant strains, allowed to recover after removal of the external stress, and finally annealed by heating above their smectic-C to isotropic transition temperatures $(T_I)$ while changes in their molecular organization were measured and analyzed. Both these materials, namely, LCE1 and LCE2, were synthesized by W. Ren, *et al.*, [42] and are important for two reasons: (1) they have remarkable strain-retention ability at room temperature and thus a potential use as shape memory materials [20], and (2) a plateau in the stress-strain curve points toward a region of soft-elastic behavior. Both of these phenomena are absent in a conventional elastomer which does not possess any micro-structural detail. The appearance of soft-elasticity and shape memory effect in LCE1 and LCE2 could only be attributed to the underlying LC microstructure.

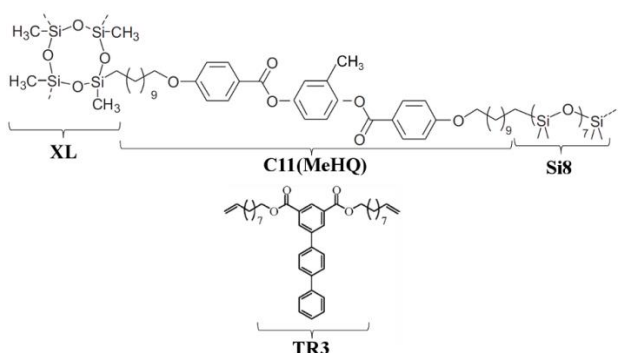

FIG. 1. Representation of elastomer LCE1 with 10 mol-% of the siloxane based cross-linker XL, rigid liquid crystalline part with floppy hydrocarbon tails $C_{11}(MeHQ)$ and elastic siloxane based chain-extender Si8. As many as four chains could attach to the cross-linker. In case of LCE2, 20 mol-% of $C_{11}(MeHQ)$ is replaced by the p-terphenyl transverse (TR3) component; adapted from [41].

In this article, we address the following important questions regarding the microscopic structure and its association with the macroscopic property of these LC elastomer networks:

1) How are the different elastic regimes in the stress-strain curve related to the structure of the elastomer and the smectic-C phase?

2) How do the underlying mesophase and the LC elastomers respond to applied strain? What is the associated relaxation dynamics, particularly in the plateau region of the stress-strain curve? Are the relaxation dynamics different in different elastic

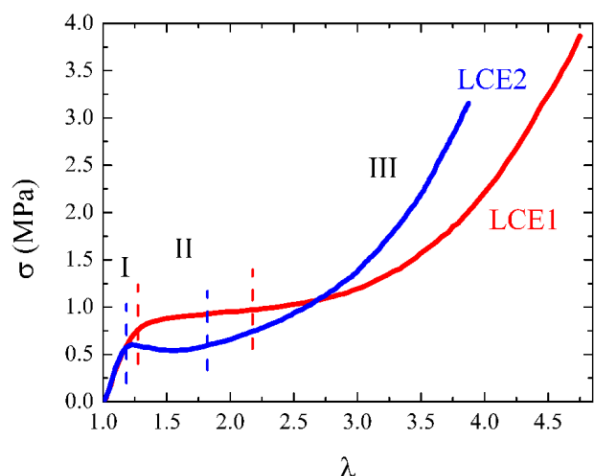

FIG. 2. Stress-strain plots of the elastomers LCE1 and LCE2. The dashed lines are the guides to the eye, separating the three major regions of the stress-strain curves for respective elastomers. Adapted with permission from [42], © (2009) John Wiley and Sons.

regimes?

3) How is the applied strain retained and released by the elastomers with time and upon heating?

In the following section, we provide details regarding chemical composition of the elastomers, their mechanical behavior, and method of application of uniaxial strain and simultaneous x-ray diffraction (XRD) measurements during a complete shape memory sequence for both the elastomers. Next, we show and discuss the results of our investigations followed by conclusions.

## II. EXPERIMENTAL PART

Chemical compositions of LCE1 and LCE2 [42,43] are shown in FIG. 1. One end of the mesogenic core $C_{11}(MeHQ)$ is connected to the cross-linker XL and the other end goes to the flexible siloxane based spacer Si8. The presence of the spacer imparts additional flexibility to the mesogenic core. XL has four open arms so that four repeat units can attach to it. Another cross-linker can attach to the open-end of Si8 and this structure is repeated inside the elastomer network. In case of LCE2, 20 mol-% of the mesogenic part $C_{11}(MeHQ)$ is replaced by the p-terphenyl transverse rod TR3 [42].

The chain extension and cross-linking reactions for preparing the elastomers were performed at room temperature. LCE1 was chosen as the parent elastomer because its cross-linking concentration is optimum for efficient liquid crystalline and elastomer network properties [41].

The stress-strain plots of both these elastomers are shown in FIG. 2. Stress ($\sigma$) is measured as force per unit initial area of cross-section of the elastomer sample (nominal stress). Strain ($\lambda$) is measured as $\lambda = 1 + \Delta L/L_o$ where, $L_o$ is the initial length and $\Delta L$ is the change in length of the elastomer specimen. The stress-strain plots are divided into three regions. Region I is the elastic region where both the materials behave similar to a conventional elastomer. Removal of the external load brings the elastomer specimens back to their initial length. Region II is the plateau region or the region of soft-elastic deformation. The stress increases slightly in this region upon application of uniaxial strain and removal of the external load only partially recovers the initial length. The polydomain to monodomain transition occurs in this region. Energy is again needed to stretch the elastomers beyond the plateau region (region III). This is the strain-retention or shape memory region. Most of the imparted secondary shape (by application of uniaxial strain) is retained by the specimens after entering region III (upon removal of the external load). In addition, LCE2 has a lower value of the threshold-strain to the plateau region, a narrower

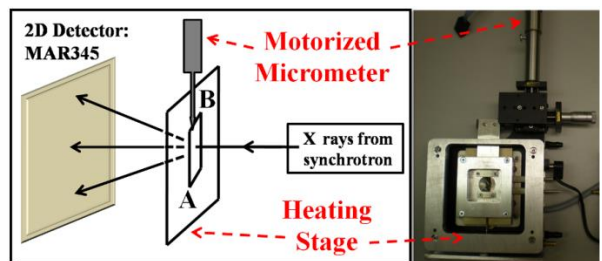

FIG. 3. Schematic of the experimental setup at 6IDB beamline at Advanced Photo Source of the Argonne National Laboratory. The customized stretching setup and accessories (on an INSTEC heating stage) are shown on the side. Both temperature of the oven and the motion of the micrometer are remotely controlled via a computer.

plateau and also, it snaps at a lower strain as compared to LCE1.

We performed the x-ray diffraction (XRD) studies at the 6IDB beamline at the Advanced Photon Source of Argonne National Laboratory. We cut small strips of elastomer specimens (6x3 $mm^2$ area and $\sim 0.3$ $mm$ thickness) and mounted the specimen(s) on the customized heating stage, between the fixed clamp A and movable clamp B (attached to the micrometer), FIG. 3. The micrometer had a precision of 122 $nm$. Wavelength of the incident radiation from synchrotron source was 0.765335 Å while the incident beam was focused to a very small cross sectional area of 100x100 $\mu m^2$. We collected the XRD patterns by using an area detector (MAR345) with 100 $\mu m$ pixel sizes. The x-ray spectrometer was calibrated using a silicon standard (NIST 640C).

We proceeded by collecting XRD data at regular intervals of time (after the changes in applied strain) during a complete shape memory sequence. The experiments on LCE1 was done with three elastomer strips, namely, S1, S2 and S3, all cut from the same specimen. We uniaxially strained S1 at an interval of $\Delta\lambda = 0.4$ from $\lambda = 1.0$ to 4.0, collected a few XRD patterns, and then moved to the next higher strain. The point of study moved upward at the introduction of a new strain and we have subsequently lowered the oven by half of that amount so that same point is studied always.

These experiments reveal that the small angle XRD pattern (SAXS) separated into four spots between $\lambda = 1.4$ to $1.8$. We removed S1 from the stretching apparatus after reaching $\lambda = 4.0$ and kept it for shape recovery measurements after ~ 24 hours. We performed the shape recovery measurements by heating the specimen above $T_I$ at the rate of 1 °C/min. A few XRD images were collected during this process. Similar experiments were also performed on LCE2.

The elastomer strip S2 was uniaxially strained in steps of $\Delta\lambda = 1.0$ at room temperature. After reaching our target strain, $\lambda = 4.0$, we released the lower clamp A (FIG. 3) but hung a small weight (~ 0.3 g) to keep the film straight. We started collecting data within ~ 2 minutes of releasing the clamp until (~ 90 minutes) no change could be ascertained in the XRD patterns. Similar measurements were done on LCE2 for ~ 40 minutes after reaching $\lambda = 3.0$.

We repeated the experiments on S3 after six months. This time, we moved at smaller strain steps ($\Delta\lambda = 0.1$), close to the polydomain to monodomain transition region at room temperature. These measurements gave us more details of the structural changes during the polydomain to monodomain transition in the parent elastomer LCE1. We used the FIT2D software package [44] to perform background subtraction, and to generate the $2\theta$- and $\chi$-scans from the XRD data.

### III. RESULTS AND DISCUSSION

FIG. 4 shows representative XRD patterns for a number of strains taken approximately twenty minutes after imparting the strain on the parent elastomer LCE1. At each strain, we observe three sets of reflections. The innermost reflection is shown on an expanded scale on the right hand side at each strain. The two wide angle (WAXS) reflections correspond to the length-scale of ~ 4.2 Å and ~ 7.2 Å and arise respectively from the hydrocarbon and siloxane segments of the elastomer which do not mix well [45] at the molecular level. The smaller dimension arises from the lateral separation between the hydrocarbon mesogenic components while a similar separation between the siloxane parts gives rise to the 7.2 Å reflections. The diffused nature of the consonant WAXS reflection points toward a random orientation of the siloxane segments. Only after $\lambda \gtrsim 1.7$ do they gain noticeable ordering, as reflected in circular reflections transforming into arc-like reflections. The smallest angle reflections (at ~ 46 Å) correspond to the smectic layer thickness. The second harmonic peaks are also present at all strains (FIG. 4: more noticeable at higher strains), originating from well-developed smectic density wave in this elastomer.

#### 1. Polydomain to monodomain transition by uniaxial strain

In this section, we shall see how the LC microstructure gradually develops from a random orientation at the state of no-deformation toward a chevron-like optically monodomain state at high strains. We shall focus on results from LCE1 but LCE2 also show similar behavior.

The measurements were started at $\lambda = 1.0$, $i.e.$, with no strain applied to the crosslinked polymer network. All the reflections remained as diffused rings, FIG. 4. At a small strain ($\lambda \sim 1.2$), the SAXS reflections concentrated into two broad, arc-like reflections perpendicular to the stretch direction. The WAXS reflections remained almost uniformly-diffused rings with very little modulation perpendicular to the stretch direction, FIG. 4. At larger strains, the SAXS reflections were widespread in the azimuthal direction, finally splitting into four spots at $\lambda = 1.7$. At the same time, the WAXS reflections corresponding to the hydrocarbon parts became more concentrated perpendicular to the stretch direction. The wide-angle siloxane ring also became marginally more intense perpendicular to the stretch direction, ultimately separating into two vertical arc-like reflections. It is to be noted that at low strains, the SAXS reflections were perpendicular to the stretch direction while the WAXS reflections remained more or less uniform rings. With increasing strain, the SAXS peaks

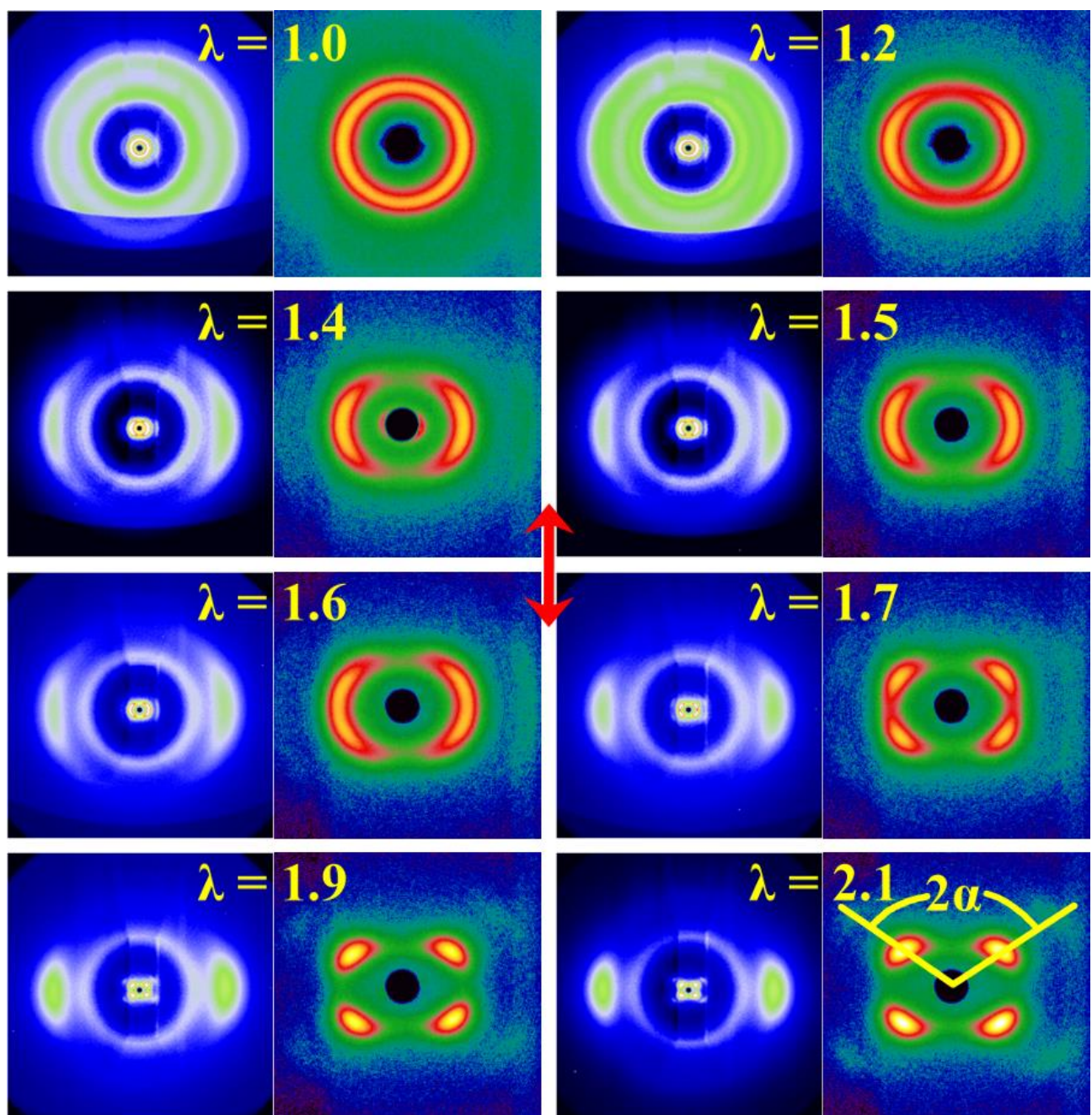

FIG. 4. LCE1: Representative XRD patterns (left images at each strain) and zoomed-in small-angle (SAXS) patterns recorded ~ 20 min after the application of the corresponding strain (marked on the patterns) applied in vertical direction. The left-side images at each strain show the effect of shadowing (at lower strains) from parts of the stretching setup and also the shadow of the elastomer film.

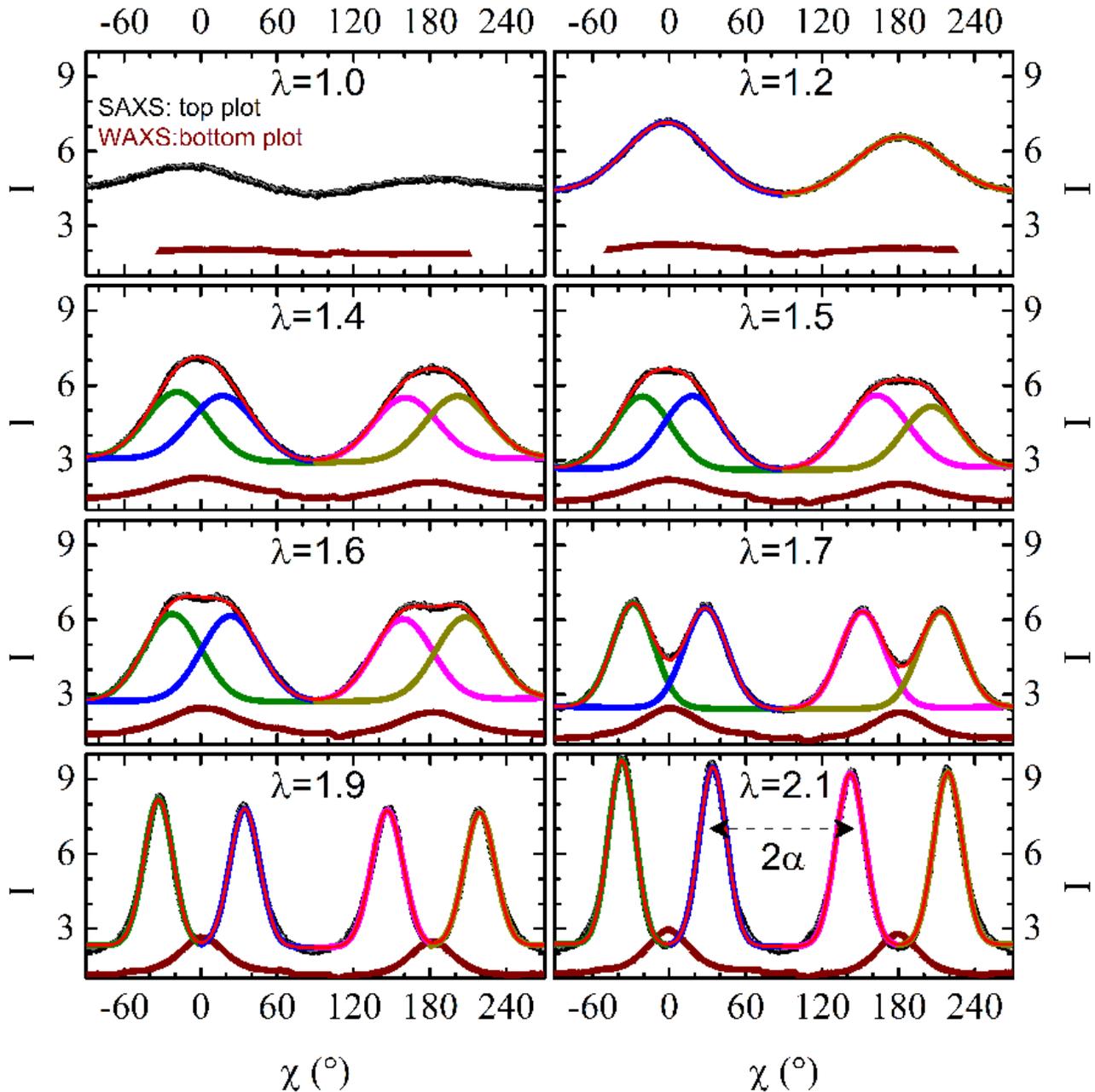

FIG. 5. LCE1: Azimuthal intensity I (arbitrary units) profiles of both SAXS and WAXS reflections in FIG. 4. The SAXS profiles correspond to the smectic-C layers and the WAXS profiles correspond to the lateral dimension of the mesogens. The negligible modulation of the SAXS peak at $\lambda = 1.0$ is due to a small strain induced during mounting this particular specimen. The Gaussian Peak functions are fitted to calculate the angle $\alpha$. At $\lambda = 1.2$, the SAXS peaks are centered around 0° and 180° and WAXS peaks are broad. At $\lambda = 1.5$, the SAXS reflections are flat at the top, implying a possible superposition of two or more peaks. Four SAXS peaks eventually emerge. The gradual separation of SAXS peaks is clear at $\lambda \geq 1.6$. The WAXS peaks continue to gradually sharpen with increasing strain at the position of separation of the SAXS peaks.

gradually became oblique to both the WAXS peaks and the stretch direction.

To quantitatively determine the degree of orientation from XRD results, we plotted their azimuthal intensity distribution in FIG. 5. The azimuthal distribution of the SAXS reflections give information about the formation of smectic-C microstructure while the azimuthal distribution of the WAXS hydrocarbon and siloxane reflections depends on how well oriented these components are. The slight modulation of SAXS reflection at $\lambda = 1.0$ is likely coming from a small strain induced during mounting this particular specimen. The azimuthal intensity distributions could be considered uniformly diffused which is the characteristic of a predominantly polydomain nature of the network, or a powder sample. At $\lambda = 1.2$, the WAXS reflection remain more or less uniform. The SAXS reflections are fit with two Gaussian line-shapes and the peak positions are found to be around 0° and 180°. This means the orientational distribution of the mesogens forming the layers is uniform in the plane perpendicular to the x rays while the smectic layers are formed parallel to the stretch direction. This can only be possible if the smectic layers are predominantly vertical, with the layer-normals distributed uniformly in the horizontal plane. At higher strains, the WAXS reflections peak around 0° and 180° which tells us that the mesogenic parts are eventually becoming parallel to the stretch direction. This is accompanied by the SAXS reflections first becoming broad (for strains $\lambda = 1.4$ and 1.5), and then beginning to split around the peak positions of the WAXS reflections. For strain $\lambda > 1.2$, the SAXS reflections are well fitted to four Gaussian functions which determine the positions of the four-spot reflections. The splitting of the SAXS reflections into four-spots in oblique direction to the WAXS reflections indicates the presence of the smectic-C mesophase in chevron-like configuration. Evidently, the mesogens have become parallel to the stretch direction and the smectic layers are tilted with respect to them, forming a chevron-like arrangement. The smectic layer-normals also incline in an azimuthally degenerate manner about the stretch direction and establish a chevron-like microstructure in the planes containing the stretch direction.

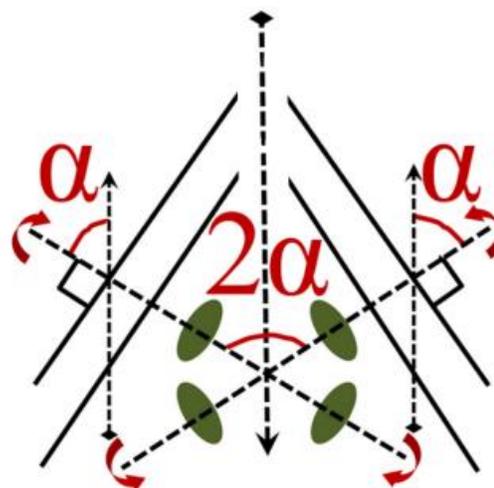

FIG. 6. Definition of the angle α and its relationship to the liquid crystal microstructure. The vertical arrows signify the direction of applied strain. The green blobs are the schematic representation for the SAXS patterns originating from these two sets of layers. The curved arrows mark the direction in which the smectic layers will rotate in response to the applied strain.

The separation 2α between the SAXS reflections in FIG. 4 is related to the chevron-like microstructure, FIG. 6. Here, α is the angle between the smectic layer-normals and the stretch direction or the direction in which long-axis of the mesogens are eventually aligned, FIG. 6. Usually, in a monomer liquid crystal exhibiting the smectic-C phase, this angle α is equal to the tilt of the mesogens with respect to the smectic layer-normal [46]. This elastomer does not exhibit an untilted (smectic-A) phase which makes the calculation of molecular-tilt from layer spacing values [46] difficult.

The angle α demonstrates a gradual decrease with increasing strain, FIG. 7 (a), saturating at a value close to ~ 50° in the chevron-like monodomain state. After the chevron-like microstructure is formed, the layer spacing $d$ also becomes smaller with increasing λ, FIG. 7 (b). A declining value of $d$ indicates corresponding increments in molecular tilt. Thus, the angle α and the tilt of the mesogens operate in contrasting ways in this elastomer. This is contrary to what is commonly observed in monomer liquid crystals where the molecular-tilt and the angle α, are directly related.

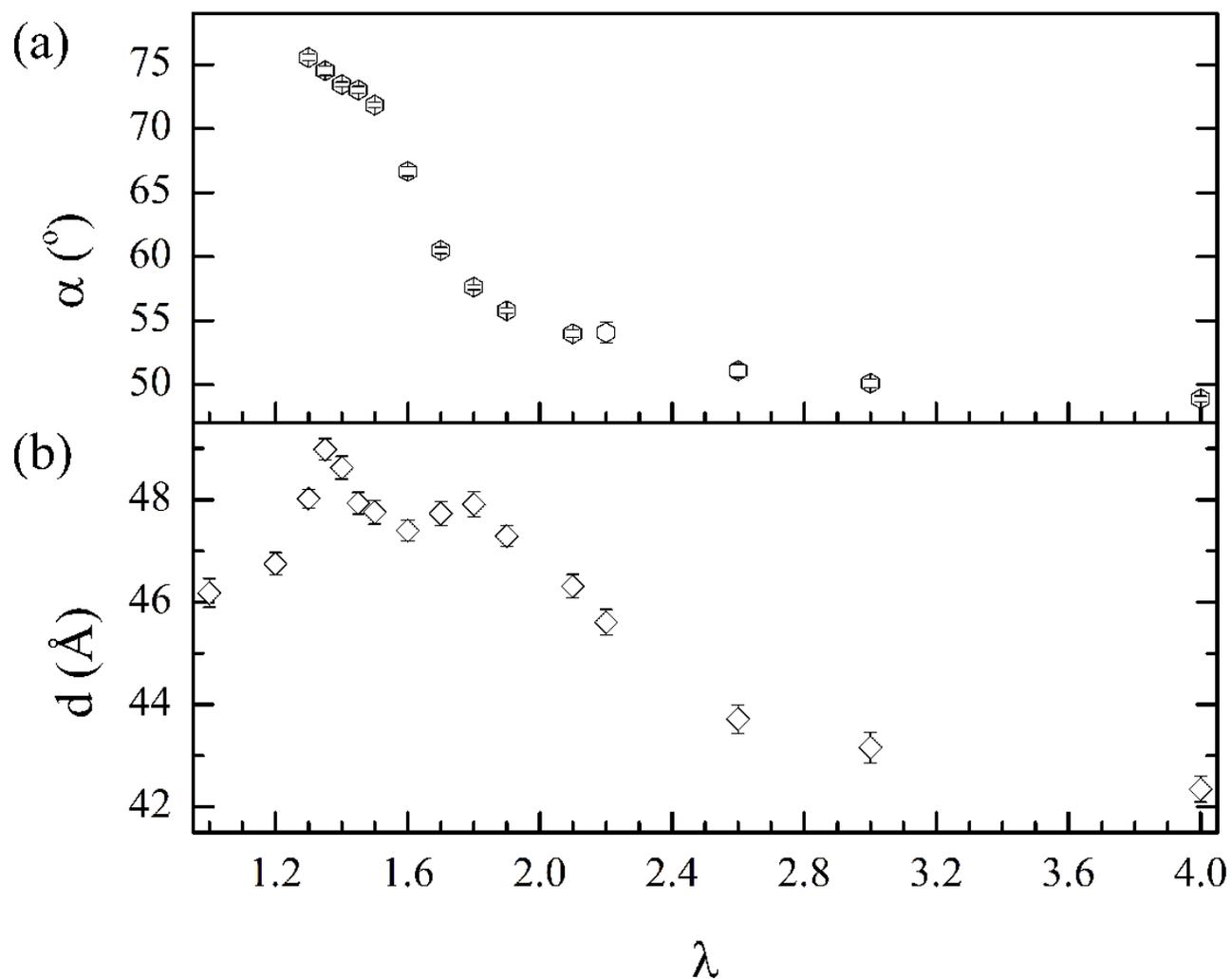

FIG. 7. LCE1: Uniaxial strain dependence of (a) the angle α and (b) the layer spacing $d$ are shown. All values are calculated after ~ 20 min of equilibration at each strain.

The global orientational order parameter $S$ of the hydrocarbon and siloxane parts is determined from the azimuthal distribution of the respective WAXS peaks, following the method of P. Davidson, *et al.*, [16]. The method is also discussed in the Appendix section. The value of $S$ (FIG. 8) calculated this way is global, because it depends on both the orientational order of the individual microdomain directors and the orientational order of the mesogens within each microdomain. The order parameter within a microdomain may be high, but on a global scale, nearly random orientational distribution of the microdomains will lead to a value of $S$ close to zero.

At low strains, a collection of disordered microdomains in the plane perpendicular to x rays gives rise to a small value of $S$, FIG. 8. With increasing strain, these domains rotate toward the stretch direction which also ensures an increase in the value of global $S$. Most of the changes in $S$ take place in the middle region which also coincides with the soft-elastic plateau (region II) in FIG. 2. The rise in $S$ conforms well to a phenomenological growth model [47], described by the equations shown in the insets of FIG. 8 (a) and (b). The model describes the changes in the orientational order $S$ whose rapid growth (across the soft-elastic plateau) is arrested as it approaches the maximum value. The rotation of the LC domains slows down beyond the value of the critical strain $\lambda \sim 1.7$. The topological constraints [9] of the network prevent a perfect alignment and the system saturates to a final orientation given by $S \sim 0.83$ and $0.4$ for the hydrocarbon and siloxane parts, respectively. The model prescribes that the secondary chevron-like monodomain be well-formed above strain $\lambda \sim 1.7$. The results shown in FIG. 4 and FIG. 5, where the splitting of the SAXS reflections is obvious for $\lambda \gtrsim 1.7$, also support this claim.

## 2. Mechanism behind polydomain to monodomain transition

Based on the results and discussions in previous section, we explain the evolution of LC microstructure in these systems, FIG. 9, as follows. The elastomer film is stretched in the vertical- or z-direction and since the volume of the elastomer is conserved [1,9], the film shrinks in both x- and y-directions, FIG. 9 (a). The x-ray beam is incident parallel to the y-direction and probes the distribution of the microstructure in the x-z plane. At zero strain, the smectic-C microdomains are distributed randomly in the 3-dimensional space inside the elastomer film. The x rays encounters a random distribution of smectic layers, mesogens and the siloxane segments, leading to uniform diffused rings at both small and wide angles, FIG. 9 (b).

Upon stretching the elastomer film slightly in the vertical z-direction, smectic layers get squeezed in the two horizontal directions (due to constancy of the volume of elastomer film). The squeezing of the layers will cause the smectic line defects [48-51] to move out to the microdomain-boundaries via a flow [52] of the smectic layers parallel to the z-direction. Uniform layers would become parallel to the stretch direction with layer-normals distributed in the horizontal x-y plane. Scattering of the x-ray beam from smectic layers (predominantly formed parallel to the stretch direction) will give rise to arc-like reflections (perpendicular to the z-axis) at small angle, FIG. 9 (c). Now this being a smectic-C system, the mesogens remain tilted with respect to the layer-normals. At low strain, the polymer chains are not appreciably influenced by the mesogens, thus the mesogens can arrange themselves randomly along the azimuthal direction inside the smectic layers maintaining a constant polar tilt with respect to the smectic layer-normals. Furthermore, the hydrocarbon linkage groups will possess little orientational preference at low strains. The system appears to have a near random distribution of hydrocarbon segments with little orientational bias giving rise to diffused wide angle rings, FIG. 9 (c).

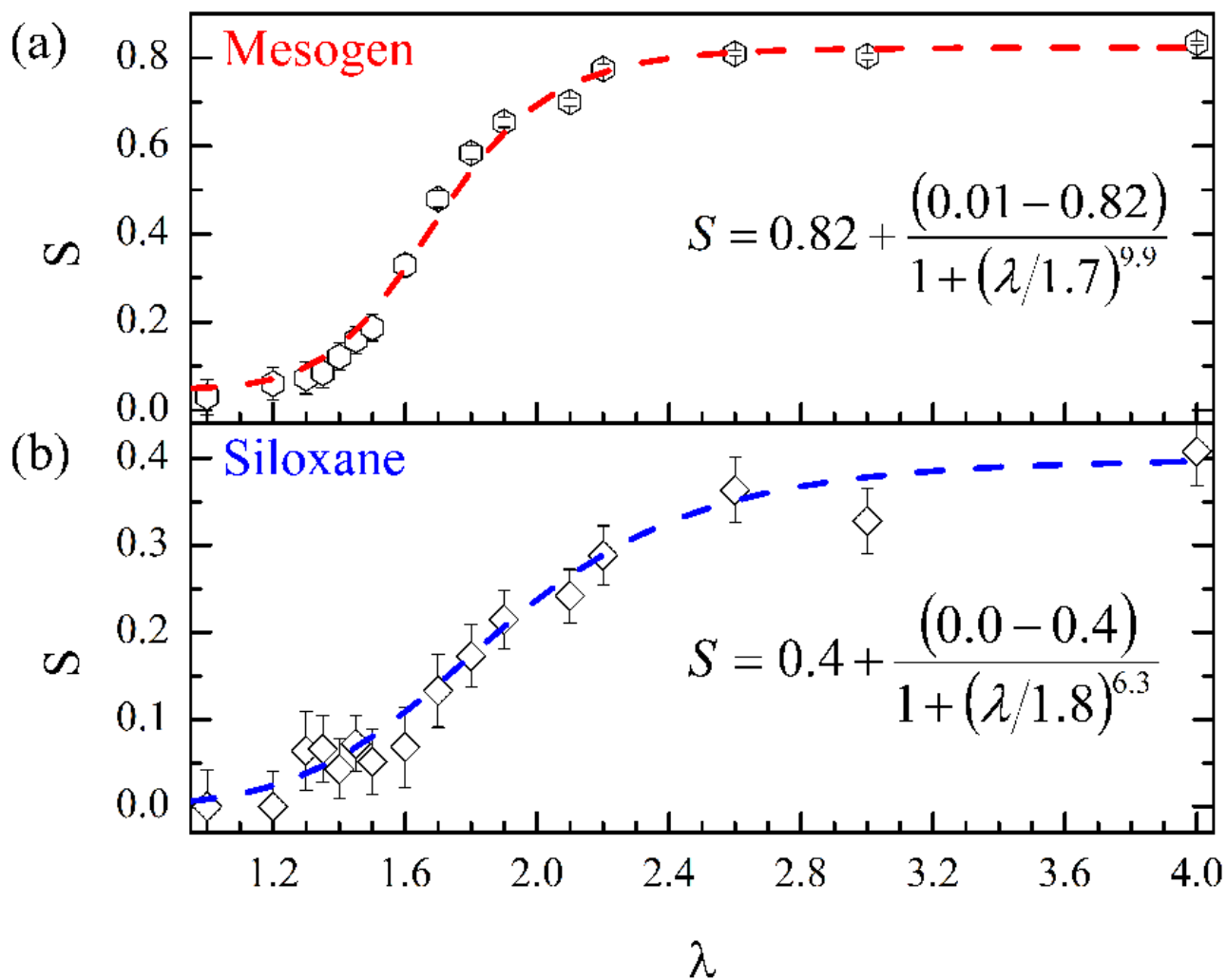

FIG. 8. LCE1: Uniaxial strain dependence of the global orientational order parameter *S* for (a) the mesogenic and (b) the siloxane segments are shown. All values are calculated after ~ 20 min of equilibration at each strain. The dotted lines are the fits according to the equations shown in the respective figures.

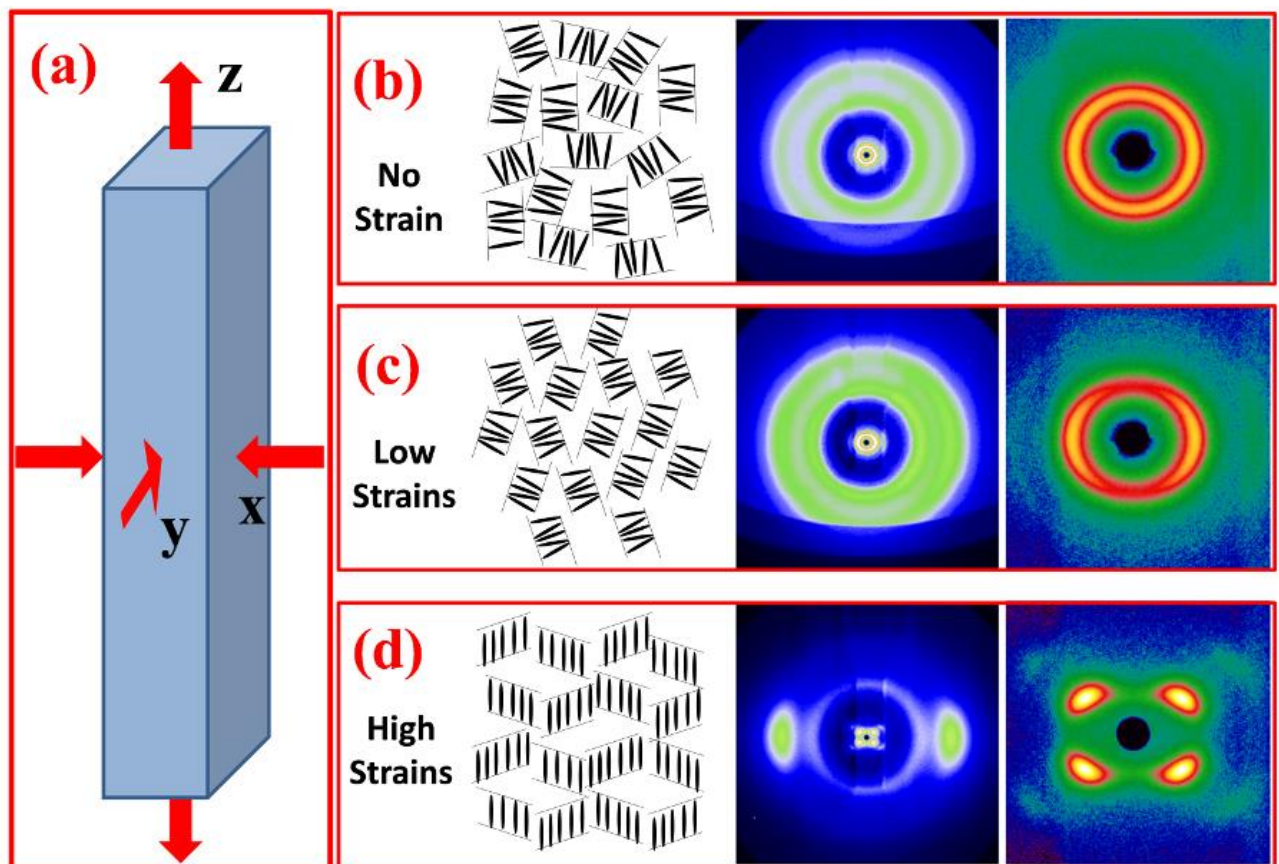

FIG. 9. (a) Schematic representation of strain applied in vertical- or z-direction to the elastomer film. The film is squeezed in both x and y-directions for conserving the volume of the elastomer. The x rays are incident parallel to the y-direction. (b) - (d) Schematic illustration of the polydomain to monodomain transition with increasing strain, associated microstructure and the corresponding all and small angle (right hand side image) XRD patterns.

At high strains, the polymer chains are stretched along the z-direction and the mesogens also become parallel to the z-direction. This is accompanied by redistribution of reflected intensity from uniform ring to two arc-like reflections at wide angle, FIG. 9 (d). The smectic-C microstructure requires a non-zero angle between the final alignment-direction of the mesogens and the smectic layer-normal. The layers rotate in a continuous fashion inside the elastomer matrix, eventually forming a 3-dimensional chevron-like microstructure at high strains and four diffraction spots at small angle, FIG. 9 (d).

To summarize, the coupling between the polymer network and LC components is very weak at low strains and the formation of LC microstructure is dominated by the in-plane flow-properties of the smectic phase [52]. At higher strains, the polymer chains strongly influence the orientation of mesogens leading to the formation of a chevron-like microstructure. The system finally attains an optically monodomain state, with the smectic layer-normals distributed in a chevron-like fashion around the stretch direction.

### 3.  Relaxation at constant strain

In initial test measurements, we collected XRD data for a long time after the application of a specific strain while the sample equilibrated. From the time dependence of the angle α, we determined the equilibration time-constant to be ~ 4 minutes in the region of high strains. In all subsequent measurements, we acquired diffraction patterns continually for at least 5 time-constants, *i.e.*, 20 - 25 minutes, to gain insight into the equilibration process and to ensure that the sample had equilibrated before acquiring the final data set and changing the strain to the higher level. These XRD images were then analyzed to obtain the values of the angle α, layer spacing $d$, and the two values of $S$ shown in FIG. 10 and FIG. 11. The red points in these figures are from XRD studies on specimen S1 where larger strain-steps ($\Delta\lambda = 0.4$) were taken to pin-point the onset and end of the polydomain to monodomain transition region. The results designated by the black points are from specimen S3 (strain-step, $\Delta\lambda = 0.1$). Both sets of results overlap very well showing the reproducibility of our results.

The angle α drops at the introduction of a new strain, then continues to drop during relaxation, and then drops again, FIG. 10 (a). Layer spacing $d$ behaves differently, FIG. 10 (b). It drops at introduction of a new strain, then increases during relaxation and then drops again. Decrease in $d$ points to an increase in the molecular-tilt which is recovered during relaxation. But the angle α does just the opposite and drops in a continuous manner. This contrasting behavior of the angle α and the molecular-tilt confirms that these two quantities in LC elastomer systems are not directly and as simply related as they are in the smectic-C phase of monomer liquid crystals. The values of the global $S$ calculated for both the mesogenic and siloxane segments show gradual increase, with jumps at the introduction of a new strain, FIG. 11. The increases at each strain appear to be related to the sudden changes in smectic ordering, reorientation, and deformation of the LC microstructure.

In the preceding section, we identified two contrasting relaxation mechanisms in LCE1. These are the flow property of the smectic layers and the elastic behavior of the polymer network and its coupling to the mesogenic parts, which dominate at the low and high strains respectively. Examining the changes in the chevron-like microstructure, or in this case, the angle α, provides an insight into these two phenomena. The relaxation of angle α is modeled using a simple exponential fit of the form:

$$y(t) = y_1 + (y_o - y_1)\exp\left[-t/\tau\right] \qquad (1)$$

where, $y_o$ and $y_1$ are the initial and saturation values of the parameter $y$ at a particular strain and $\tau$ is the relaxation time constant. A higher value of $\tau$ points toward a slow relaxation process and vice versa. Results of the fits to eqn. (1) are shown in FIG. 12 (a). These fits make a clear distinction between the relaxation behavior before and after the polydomain to monodomain transition. The values of $\tau$ calculated from these fits are plotted in FIG. 12 (b) as function of increasing strain. At small strains ($1.3 \leq \lambda < 1.7$), the relaxation of $\alpha$ is relatively slow with $\tau \sim 45$ min. As the strain is increased ($\lambda \geq 1.7$), the relaxation becomes faster and $\tau$ attains the value $\sim 5$ min. The system clearly responds differently to low and high applied strains with different values of the relaxation time constant $\tau$. In the polydomain to monodomain transition regime, the slow reformation or realignment of LC microdomains seems to be dominated by the flow properties of the smectic-C phase. At these low strains, higher flexibility of the polymer chains would allow the layers to move almost freely inside the elastomer network. The layers possibly anneal via slow movements of edge dislocations [48-51] giving rise to a higher relaxation time constant.

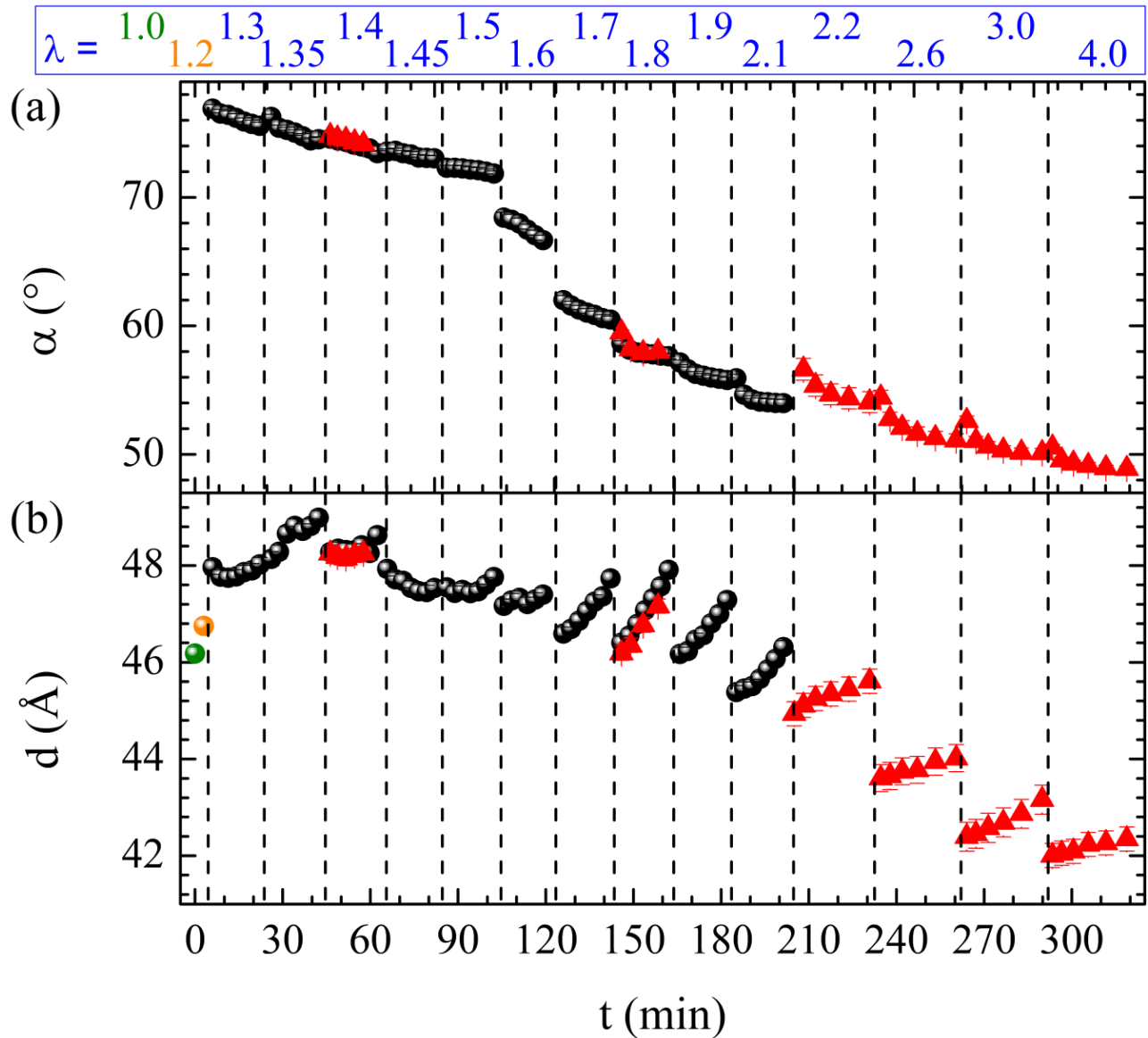

FIG. 10. LCE1: Time dependences of (a) the angle α and (b) layer spacing *d* are shown during stretching across the polydomain to monodomain transition region. The vertical dashed lines mark different strain regions with the values of strain λ shown at the top. The green and orange points correspond to λ = 1.0 and 1.2 respectively. The red- and black-points are from measurements on specimen S1 and S3 respectively. Both specimens were cut from the same elastomer sample. The results also show the reproducibility of our measurements.

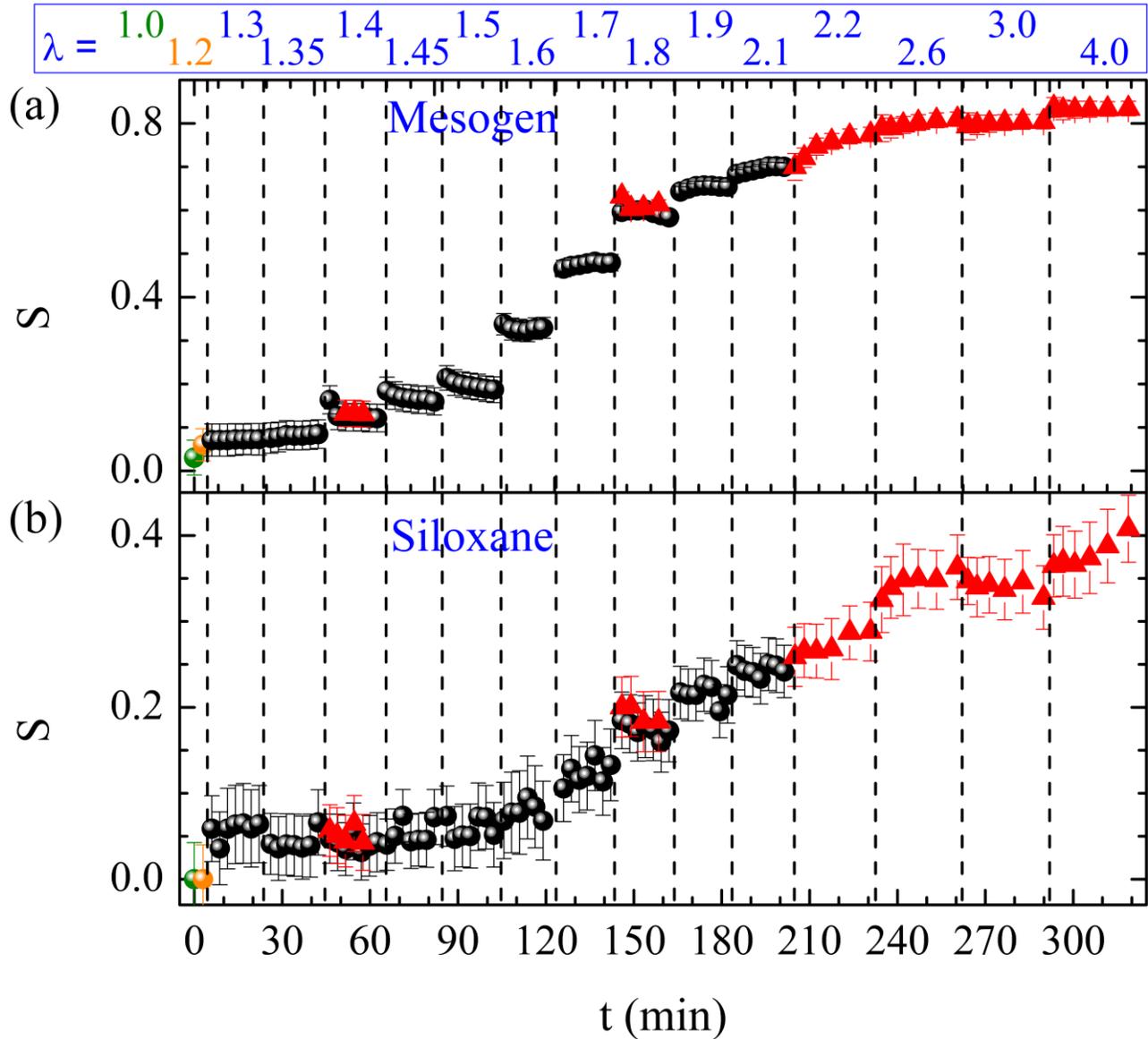

FIG. 11. LCE1: Time dependences of the global orientational order parameter $S$ for (a) the mesogenic and (b) the siloxane segments are shown during uniaxial deformation. The vertical dashed lines mark different strain regions with the values of strain $\lambda$ shown at the top. The green and orange points correspond to $\lambda = 1.0$ and 1.2 respectively. The red- and black-points are from measurements on specimen S1 and S3 respectively. Both specimens were cut from the same elastomer sample. The results also show the reproducibility of our measurements.

At strains higher than $\lambda = 1.7$, the system relaxes with an order of magnitude smaller relaxation time ($\tau \sim 4 - 8$ minutes). In this regime, the reorientation of the smectic domains is completed and the local directors of the individual microdomains, the polymer chains, and the mesogens are all pointing toward the stretch direction. This being a main-chain system, the relaxation properties would be dominated by the properties of the polymer components at high strains. This is substantiated by the fact that stress-relaxation typically occurs over a time scale of $\sim 7$ minutes (at $\lambda \sim 2.75$) [30] in an elastic network of polyisoprene rubber lacking any LC microstructure. Also, earlier stress-relaxation studies [41] on LCE1 have indicated relaxation times of the order of $\sim 4 - 5$ minutes for strains $\lambda \sim 1.5$ to 4.0. Thus, at high strains, the elastic properties of the polymer network seem to be the dominant mechanism leading to faster relaxation times.

To our knowledge, this is the first time that the relaxation time constants of LC microstructure are measured and shown to be associated with the two components of the LC elastomer systems. In other words, we have been able to separate out the role of two basic components of the LC elastomer systems and their effect on the macroscopic property of the elastomer network.

## 4. TR3 and strain retention

Previous stress-strain measurements [45] on both LCE1 and LCE2 indicate that a considerable amount of strain is retained even after the removal of the external load and the elastomer remains in this state for a very long time. Typically, more than fifty percent of the applied strain is retained in both the elastomers [45]. Deep in the third region of the stress-strain curve, *i.e.*, in the chevron-like monodomain region, the strain retention abilities of these elastomers were determined by removing the lower clamp after reaching $\lambda = 4.0$ for LCE1 and $\lambda = 3.0$ for LCE2. A negligible mass ($\sim 0.3$ g) was attached at the lower end to hold the elastomer films straight. Approximately two minutes after releasing the lower clamp, we started collecting the XRD data for $\sim 90$ minutes in the case of LCE1 and $\sim 40$ minutes in the case of LCE2. We also collected one XRD data after $\sim 24$ hours to determine the final structure after both the samples had fully equilibrated, FIG. 13 (a) and (b). The SAXS images show that chevron-like microstructure is well retained in both the elastomers for a considerable time period. The persistence of the four-spot diffraction pattern implies that the layer-normals remain oriented with respect to the stretch direction long after removal of the external stress.

To compare the strain retention abilities of LCE1 and LCE2, we calculated the values of $\alpha$ and $S$ from the XRD patterns as a function of time. For the purpose of comparing the relative changes, the values of these parameters were normalized by dividing them with their corresponding value at t = 0 min. The results are plotted in FIG. 13 (c). Both $S$ and $\alpha$ relax toward their respective final values, albeit with different relaxation time constants. Single exponential fits, to the respective data reveal that $S$ and $\alpha$ approach their final values at a faster rate in LCE2 than in LCE1.

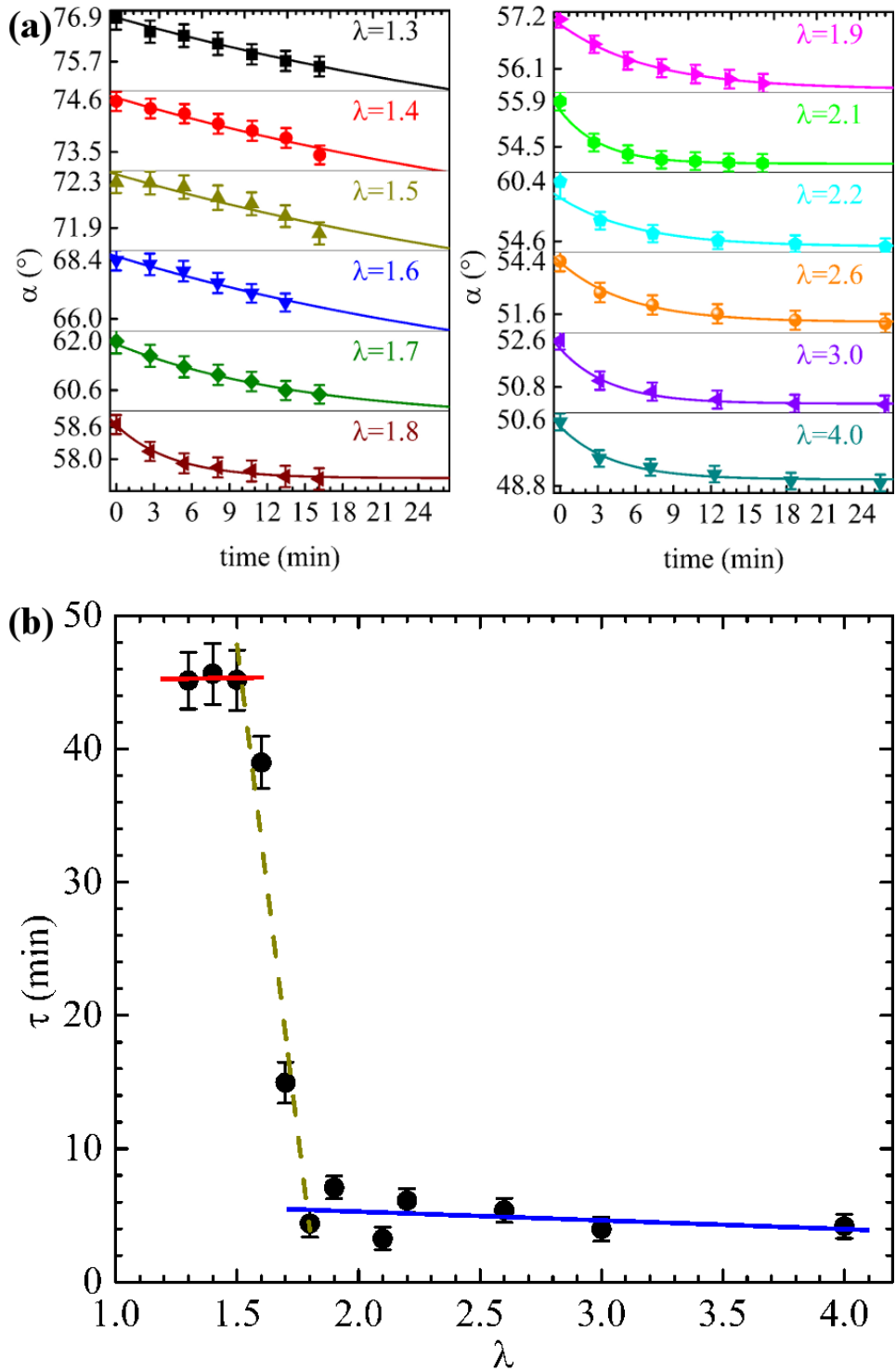

FIG. 12. (a) Relaxation behavior of the angle α at different values of strain, λ. The solid lines are the fits to the eqn. (1); (b) Plot of relaxation time constant, τ as function of strain, λ. The values are calculated from the fits of eqn. (1) to the experimental data in FIG. 12 (a). The colored lines are guides to the eye showing the transition clearly.

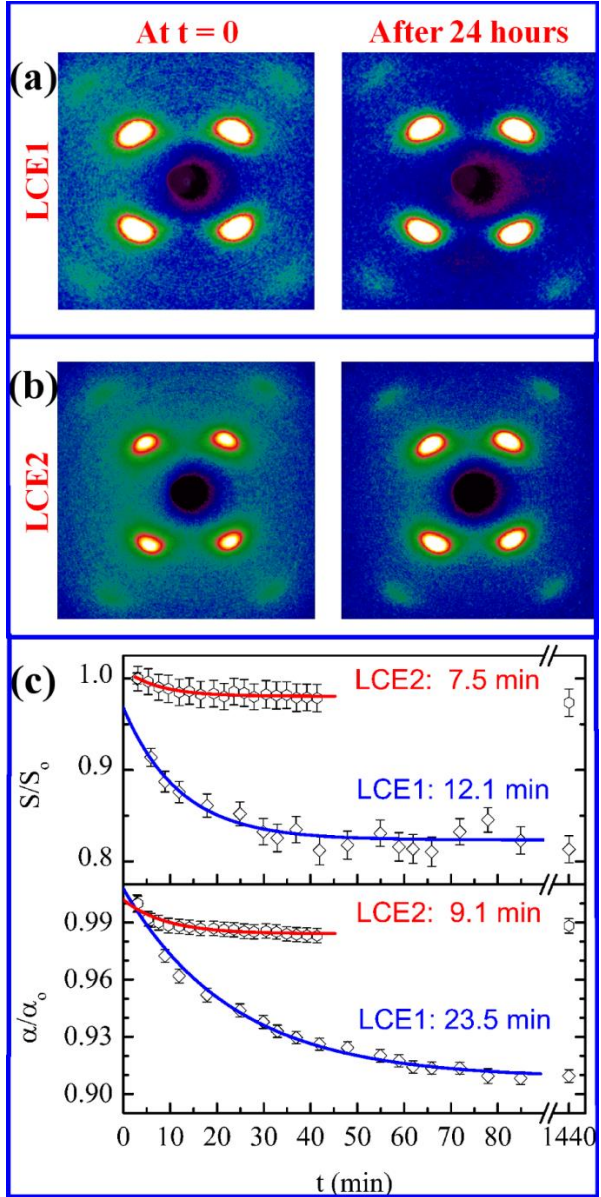

FIG. 13. (a) - (b) Small-angle XRD patterns of LCE1 and LCE2 at t ~ 0 and after t ~ 24 hours; (c) comparison of the values of $\alpha/\alpha_o$ and $S/S_o$ during strain retention experiments. Here, $\alpha_o$ and $S_o$ are the values of $\alpha$ and $S$ at t ~ 0 min. The solid lines are simple exponential fits to the data according to the eqn. (1). The values of corresponding relaxation time constants are also mentioned adjacent to the respective fits.

The relaxation time constants of $S$ are found to be ~ 7.5 minutes in LCE2 and ~ 12.1 minutes in LCE1. The time constants for relaxation of the angle $\alpha$ are measured to be ~ 9.1 minutes in LCE2 and ~ 23.5 minutes in LCE1. Clearly, the introduction of TR3 seems to enhance retention of the strain-induced orientational alignment of microdomains and the angle $\alpha$. The introduction of TR3 has thus lead to a more stable chevron-like microstructure in LCE2 and "locking" of the arrangement of the domain distribution with respect to the elastomer network.

## 5. Effect of TR3 on thermal shape recovery

The secondary shape induced by the application of strain is stable for a long period of time. It is thought [53] that the presence of unfolded hairpins in the stretched state and trapping of the siloxane based cross-linkers in the siloxane rich regions of the elastomers lead to the stability of the secondary shape. As already shown in FIG. 13 (c), the introduction of TR3 aids LCE2 in retaining more than 95% of the macroscopic $S$ compared to LCE1 which retains only about 80%. The initial polydomain state of such crosslinked networks could be restored by either swelling the network in acetone or by raising the sample temperature above $T_I$ [41]. We performed thermally driven shape recovery where the elastomer films were gradually heated above their respective $T_I$ (~ 96.1 °C for LCE1 and ~ 70.6 °C for LCE2).

FIG. 14 (a) shows the SAXS patterns during the chevron-like monodomain formation by uniaxial stretching in LCE2. Subsequent restoration of random distribution of the smectic-C domains by heating is shown in FIG. 14 (b). During the polydomain to monodomain transition, the SAXS peaks appear first as two crescents *perpendicular* to the stretch direction, FIG. 14 (a). They then split into two pairs of peaks at high strain forming a smectic-C chevron-like monodomain. The presence of the second order smectic layer peaks at higher strains establishes highly-condensed smectic density wave [54] in the system. As the thermal recovery process begins, the intensity of the four SAXS peaks diminishes with temperature, then they merge into each other forming arc-like reflections in the direction *parallel* to applied strain, and finally transform into an isotropic ring at temperatures above $T_I$, FIG. 14 (b). This behavior is observed in both LCE1 and LCE2, the extent of the change being more apparent in LCE2.

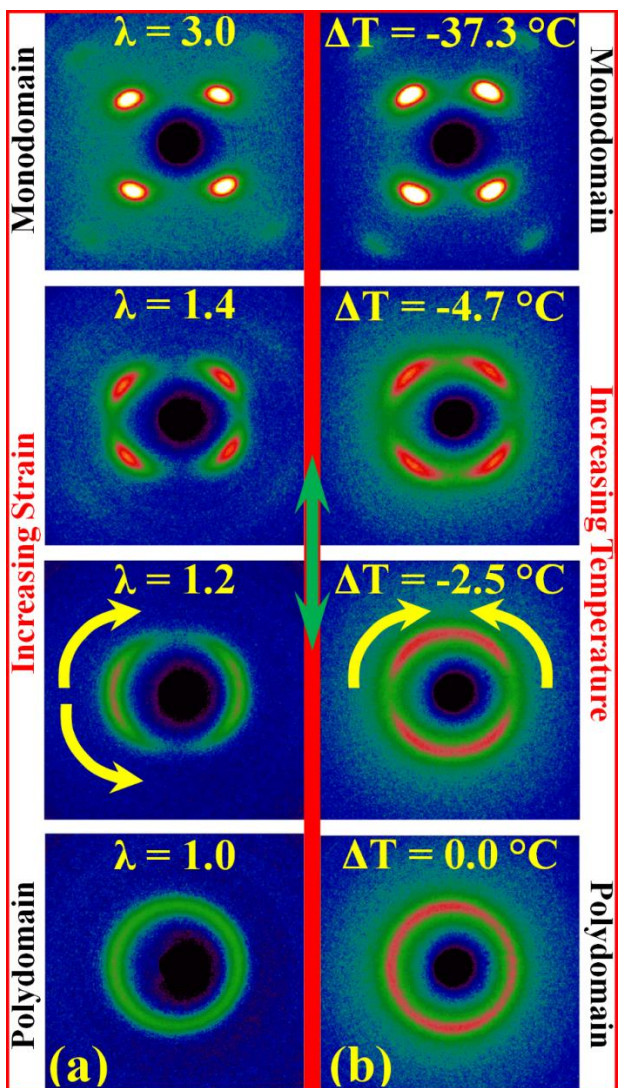

FIG. 14. LCE2: (a) Small-angle XRD patterns during strain induced polydomain to monodomain transition (strain increase from bottom to top). The curved arrows show the direction in which SAXS reflections separate into four-spot reflections. The green arrow in the middle shows the stretch direction. (b) SAXS patterns during thermal shape recovery (temperature increase from top to bottom). The curved arrows show the direction in which SAXS reflections merge into uniform rings.

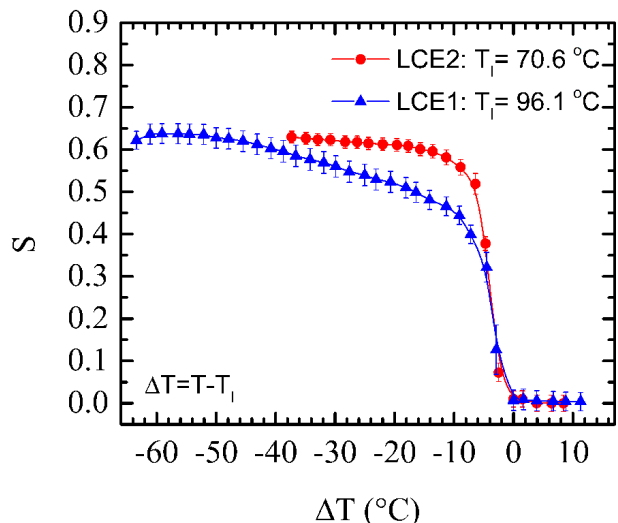

FIG. 15. Plots of global orientational order parameter $S$ during thermally driven shape recovery process of LCE1 and LCE2. The values of smectic-C to isotropic transition temperature, $T_I$ for both the elastomers are mentioned in the figure. LCE2 has a lower $T_I$ so the temperature range covered for it is narrower as the experimental setup did not provide access to below room temperature. The solid curves in the plot for $S$ serve as guide to the eye.

During thermal shape recovery process, the layers tend to form perpendicular to the initial stretch direction with an azimuthally degenerate distribution of the mesogens in the plane of the smectic layers. This is somewhat different than what we observed during the polydomain to monodomain transition by uniaxial strain where the layers were formed parallel to the stretch direction. This contrasting behavior during shape recovery can be explained by considering orientational preference of the mesogens parallel to the stretch direction in the chevron-like monodomain state. The chevrons re-merge, forming poorly defined layers that are statistically perpendicular to the stretch direction, until the elastomers enter the isotropic phase above $T_I$.

The disappearance of the second harmonic reflections points toward a partial loss of the macroscopic smectic ordering with heating. At and above $T_I$, the smectic-C mesophase transforms into the isotropic phase and all reflections turn into uniform rings. We determined the value of $T_I$ from SAXS results and it is found to be ~ 10 °C lower than the value measured by differential scanning calorimetry (DSC) [42]. The DSC curve is broad [45] and previous stress-strain measurements [53] have related this shape recovery temperature $T_I$ to the onset of the isotropic phase in these materials. Our results are consistent with that observation.

FIG. 15 shows the changes in global orientational order parameter $S$ during the thermal shape recovery for the two elastomers. With rising temperature, $S$ for LCE1 and LCE2 show a smooth transition to the isotropic value ($S = 0$). The value of $S$ in LCE2 falls to zero more abruptly across the first order transition [55,56] at $T_I$ than in LCE1. The gradual change of $S$ in these elastomer systems arises from randomization of LC microdomains at high temperatures. The relatively sharper drop in $S$ across the transition (in case of LCE2, FIG. 15) again points toward a greater tendency for strain-retention in LCE2 due to the incorporation of TR3.

## IV. CONCLUSIONS

We have determined the structural changes that occur during the application of strain, relaxation at constant strain, recovery after removal of the external stress, and thermal shape recovery processes of two main-chain smectic-C elastomers using x-ray diffraction. The stress-strain curves of these materials showed a plateau which is the region of soft-elastic deformations. We have identified the relaxation dynamics in these systems pertaining to the polydomain to monodomain transition in this regime of the stress-strain curve. Initial random director and smectic layer distribution of smectic-C microdomains inside the elastomer network is found to gradually change into a chevron-like optically monodomain configuration with the application of strain. The strain-induced alignment and re-orientation of liquid crystal microstructure proceeds slowly ($\tau \sim 45$ minutes) at low strains. The time-constant $\tau$ decreases by at least an order of magnitude ($\tau \sim 5$ minutes) as the polydomain to monodomain transition is completed at higher strains. We attribute the slow relaxation process at low strains to flow properties [52] of the liquid crystal layers embedded in the elastomer network. At higher strains, the elastic response of the polymer component becomes dominant, leading to a faster relaxation process for the liquid crystal microstructure.

The value of the global orientational order parameter $S$ is initially close to zero due to the misaligned microdomains. Across the polydomain to monodomain transition, the domains attain a preferential smectic layer and director alignment parallel to the stretch direction and the value of $S$ rapidly increases from $\sim 0.15$ to $\sim 0.83$. An appreciable increase in $S$ for the siloxane parts is noticeable only after the formation of chevron-like monodomain is completed and reaches a maximum value of $\sim 0.4$ for $\lambda \sim 4.0$ for the parent elastomer LCE1.

In the final high-strain state, various components of the system, *i.e.*, the polymer chains, the mesogens, and the local microdomain directors, all align parallel to the stretch direction. The layers form oblique to the stretch direction conforming to the structural property of the smectic-C phase. This chevron-like monodomain structure is enhanced at high strains and both elastomers are found to be "locked-in" this secondary state even after removal of the external stress.

The chevron-like microstructure is found to remain well-formed even after a full day of relaxation. The relaxation of the chevron-structure toward the equilibrium state is found to be faster in the second elastomer LCE2, most likely due to presence of the transverse component (TR3) in its main-chain. Thermal energy, as an external stimulus, gradually disrupts the smectic order and both elastomers recover their initial polydomain state above their respective smectic-C to isotropic transition temperatures. A preference for the orientation of the smectic layer-normals toward the stretch direction persists in the chevron-like monodomain state until a random distribution of the liquid crystal microdomains is achieved. The introduction of the TR3 component enhances the strain-retention ability in LCE2.

### ACKNOWLEDGEMENTS

This research used resources of the Advanced Photon Source, a U.S. Department of Energy (DOE) Office of Science User Facility operated for the DOE Office of Science by Argonne National Laboratory under Contract No. DE-AC02-06CH11357. The data was collected at the X-ray Science Division 6-ID-B beamline at the Advanced Photon Source, Argonne National Laboratory. SD duly acknowledges the Physics Department at Kent State University (KSU) and prof. Satyendra Kumar for the financial support during his stint as a graduate student there. SD is grateful to prof. Satyendra Kumar for insightful discussions and allowing him to use his laboratory. His critical inputs are duly appreciated. Dr. Dena Agra-Kooijman is also acknowledged for her contributions, wherever appropriate. Authors are grateful to prof. Jonathan Selinger for useful discussions. Dr. Alan Baldwin is acknowledged for his invaluable support regarding the use of the motorized micrometer, a vital piece of equipment used in this study. Mr. Wade Aldhizer's support in the KSU physics machine shop is also acknowledged.

## APPENDIX: CALCULATION OF $S$, THE ORIENTATIONAL ORDER PARAMETER

The orientational order parameter, $S$ for the siloxane and hydrocarbon parts are calculated from the wide-angle scattering data following the method of P. Davidson, $et$ $al.$, [57]:

$$S = \frac{1}{2}[3\langle \cos^2 \beta \rangle - 1] \qquad (2)$$

Here, $\beta$ is the angle made by the molecular segment with respect to the stretch direction or the macroscopic director. It is possible to parameterize the WAXS reflections with respect to the azimuthal angle $\chi$ on the detector plane [57]:

$$I(\chi) = [I_o + a\chi] + K \cdot \frac{\exp(b\cos^2 \chi)}{4\pi\sqrt{b}\cos\chi}\left[\frac{\int_0^{\sqrt{b}\cos\chi}\exp(-x^2)dx}{\int_0^1 \exp(bx^2)dx}\right] \qquad (3)$$

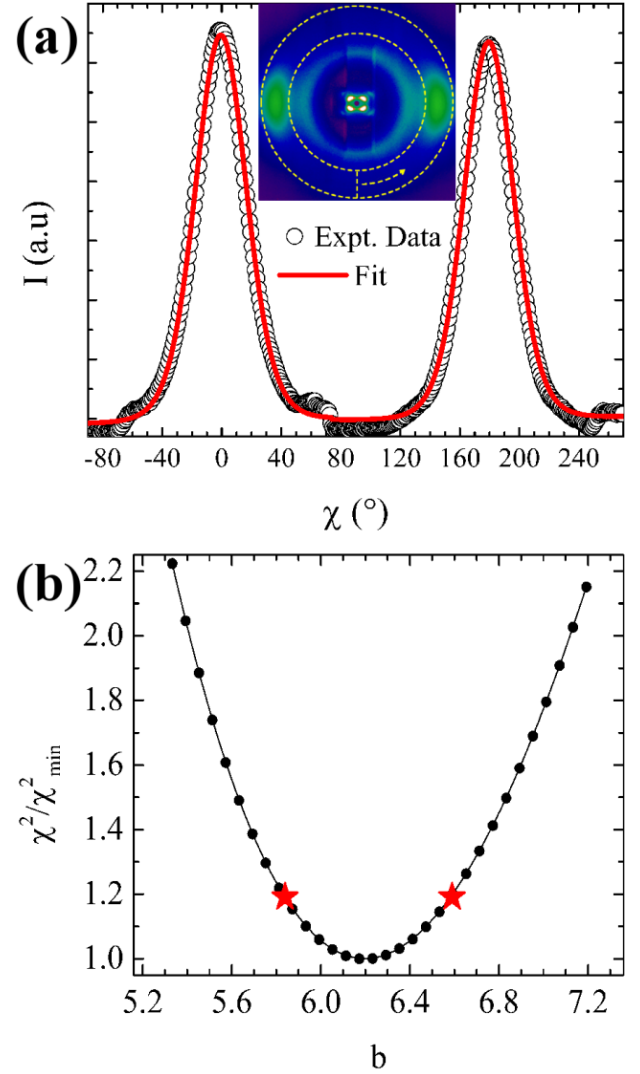

FIG. 16. (a) Plot of $I$ vs. $\chi$ for the wide-angle area (between the yellow dotted circles) of the representative XRD pattern in the inset. The solid line is the fit to the experimental data according to the eqn. (3) which gives the value of the fit parameter $b$; (b) $\chi^2$, which is a measure of goodness of fit [58], is plotted against the fit-parameter $b$. The values of $\chi^2$ are normalized with respect to the best fit value. The ★'s correspond to the 95% confidence limits determined by the $F$-test: $F(356,356) \approx 1.191$. From this plot, error in the value of the fit-parameter $b$ is calculated.

A numerical fit of eqn. (3) is performed on the '$I$ vs. $\chi$' profiles of the wide-angle reflections, FIG. 16 (a). The best fit values of the fit-parameters $I_o$, $a$, $K$ and $b$ are determined by minimizing the value of $\chi^2$, which is a measure of goodness of fit [58].

Using the value of the fit-parameter $b$, the average $<\cos^2\beta>$ is first calculated [29]:

$$\left\langle \cos^2\beta \right\rangle = \frac{\int_0^1 x^2 \exp\left(bx^2\right)dx}{\int_0^1 \exp\left(bx^2\right)dx}; \text{ where, } x = \cos\beta$$

and then by using eqn. (2), the value of $S$ is obtained.

Only the fit-parameter $b$ contributes in determining the value of $S$. Uncertainties in the value of $b$ are calculated by fixing all the other parameters at the best-fit value and then changing the value of $b$ on either side of the best-fit value [59] while the data are refitted according to the eqn. (3). The values of $\chi^2$, calculated from these fits, are plotted against the corresponding value of $b$, FIG. 16 (b). These values form a parabola, with minimum at the best-fit value of $b$. Next, the $F$-test [58] is employed to determine the probable uncertainty in the value of $b$ within the 95% confidence level. Then, using the method of error-propagation [58], error in the measured quantity $S$ is calculated.